\documentstyle[psfig]{article}
\begin{document}
\Large{
{\bf
\noindent
THE PHOTOMETRIC AND SPECTRAL INVESTIGATION OF CI CAMELOPARDALIS,
AN X-RAY TRANSIENT AND B[e] STAR\\[1mm]

\noindent
E.A.Barsukova$^1$, N.V.Borisov$^1$,V.P.Goranskij$^2$, V.M.Lyuty$^2$,
and N.V.Metlova$^2$.}\\[1mm]

\noindent
$^1$ Special Astrophysical Observatory, Russian Academy of Sciences,
Nizhny Arkhys, Karachaevo-Cherkesia, Russia \\[1mm]
$^2$ Sternberg Astronomical Institute, Moscow University,
Russia \\[1mm]


\noindent
We combined the results of  $UBVR$ photometry of CI Cam taken at Sternberg
Astronomical Institute
in 1998--2001, and moderate resolution spectroscopy taken at Special Astrophysical
Observatory during  the same time period.
Photometry as well as fluxes of Balmer emissions and of some Fe~II
emission lines of CI Cam in quiet state reveal a cyclic variation
with the period of $1100\pm50^d$.
The variation like this may be due to an orbital motion in a wide pair with
a giant star companion that exhibits the reflection effect on its side faced to a
compact companion.

The $V$-band photometry also confirms the pre-outburst 11.7 day period found
by Miroshnichenko earlier, but with a lower amplitude of 3 per cent.
The possibility of identity of this photometric period with the
period of jet's rotation in the VLA radio map of the object CI Cam was investigated.
The radio map modelling reveals the inclination of the jet rotation axis to
the line of sight, $i = 35-40^o$, the angle between the rotation axis
and the direction of ejection of the jet, $\theta = 7-10^o$, and  jet's
spatial velocity of 0.23--0.26$c$.

Equivalent widths and fluxes of various spectral lines show
different amplitudes of changes during the outburst, and essentially
distinct behaviour in quiescence. Five types of such behaviour were revealed,
that indicates the strong stratification of a gas and dust
envelope round the system . The time lag of strengthening of 50--250$^d$
in the forbidden line of
nitrogen [N~II] was found relatively to the X-ray
outburst maximum.
}

\newpage
\Large{
\section*{Introduction}

CI Cam (MWC 84, KPD 0415+5552, MW 143, LS~V +55$^o$16,
IRAS 04156+5552,\ XTE~J0421+560;\ $4^h19^m42^s.11,\ +55^o59'57''.7$,\ 2000)
is a well-known B[e] star, $V=11^m.6$, an X-ray system, and a microquasar.
Stars having the following spectral features are usually attributed to
B[e] star class [1]:
(1) bright hydrogen and helium emissions above the blue continuum;
(2) numerous bright lines of Fe~II and other metals, mainly at
low ionization levels  and with low excitation potential;
(3) forbidden emission lines of metals and of other elements in optical
spectrum, e. g. [Fe~II], [O~I];
(4) strong infrared excess due to radiation of hot circumstellar dust.
\hbox{CI Cam} does have these features in its spectrum.  Since 1932, it is known
 as a B star with H$_\alpha$, He~I and  Fe~II emissions [2,3].
The spectrum of the star was thoroughly studied  in [4,5].
The results of broad-band photometry in  optical and infrared
diapasons are presented in [6]. The star showed a significant light
variability with the amplitude up to $0^m.4$ in the $V$ band, and therefore was
included into General Catalogue of Variable Stars. The photometric
11.7 day period was determined in [5,7] based on data [6],
the amplitude of periodic component being $0^m.3$~$V$.
The sum of two spectra  B0V + G8II [7], or B0V + K0II [5]
fits the spectral energy distribution of this system.
The observation of absorption lines from a cool star is even reported in [5].
Thus, CI Cam resembles  symbiotic variable stars, which have properties overlapping
largely with the B[e]-stars ones. The  light of star is considerably
absorbed. The significant part of this absorption has a circumstellar
origin. The star's distance is of 1 kpc [8].

In 1998, April the star underwent a powerful X-ray outburst, its
2--12 KeV flux reached 2 Crabs at maximum on April 1, $0^h57^m$UT
(JD 2450904.54). The X-ray flux raise to maximum lasted for a few hours [9].
The posterior
fading was violent, too, the first two days the flux decreased
with the parameter $\tau_e = 0^d.56$ ($\tau_e$ is the time of fading
by $e$ times). The rate of fading on the 4-th day was in accordance
with $\tau_e = 2^d.34$ (slowing down of fading). The X-ray spectrum
was soft relative to X-ray novae, and did not expand in the
energy range above 60 KeV [10]. The following emissions were seen: the
K line of Fe~XXV--Fe~XXVI at 6.4--6.9 KeV, S~XV--S~XVI at 2.45--2.62
KeV, and the L line of  Si~XIII-Si~XIV at 1.86--2.01 KeV [11].

Optical observations were carried out with some delay relatively
to the X-ray maximum, which is necessary for identification.
The greatest brightness registered in the outburst was 7.1 $R$ [12].
Considerable brightening
of emission lines occured in optical and infrared spectra [13,14], and
the strong He II 4686 \AA\ emission appeared [13]. H and He emissions
had wide pedestals showing the expansion velocity of 1200 km/s
in the star envelope [15]. The contribution of emission lines to
photometric bands increased. Radio source was detected by
April 1, and got maximum of 120 mJy at 1.4 GHz on April 3.83~UT.
The radio flux curves, as well as those of X-ray, optical, and
infrared fluxes versus time were published in [16]. The time delay
of the outburst maximum depending on the increasing of radiation wavelength
was noticeable in radio. The rate of optical brightness decline was
$\tau_e = 3^d.4\pm0^d.4$ since April 3 to April 19, 1998.

The radio source is a syncrotron source by nature. On April 5 it was resolved
into a central core and opposite directed jets, which got later
S-shaped appearence similar to the radio jets of SS~433. The radio
map was published in [17]. The proper motion of each jet was 26
milliarcseconds per day, or 0.15$c$ assuming the distance of 1 kpc
[18]. A structure of jets is symmertic, and they are corkscrewing along
a conic surface. As a galactic stellar system with jets, CI Cam
entered the class of objects called "microquasars".

The X-ray flux had declined down to the detectable level in 10--15 days
after peak of the outburst. An increased brightness in the
optical range was observed till the end of observing season at
JD 2450941 ($37^d$), but it got the low level in the beginning of the
next season at JD 2451051 having later the tendency of brightening
[16]. At the same time the radio flux was measurable during all
the next season JD 2451020-51300 with tendency of decline [16].
The object was detected in X-rays with BeppoSAX satellite in the
quiescence using long exposures with soft spectrum on JD 2451425
[19], or hard spectrum on JD 2451445 [20], or not detected at all.
These observations reveal the variability of the object at least by
order of magnitude. Spectral study of CI Cam in December 1998 and
January 1999 [19] showed no spectral changes
during that time period and similarity to the pre-outburst spectrum [4].

This paper presents the results of our photometric and spectral
monitoring of CI Cam in quiescence in 1998--2001, and the comparison
of photometrically calibrated spectra taken in outburst of 1998 and
in quiescence.

\section*{Photometric observations}

Photometric observations began on 1998 April 3 as soon as
we had received the information about the outburst. The observations were carried
out with SBIG ST-6 CCD  at 60-cm Zeiss reflector of the Sternberg Institute's
Crimean Laboratory. The monitoring was continued during 3 years by
N.V. Metlova with the same telescope and one-channel $UBV$-photometer
constructed by V.M.~Lyuty. The comparison star GSC 3723.54 ($V=10^m.401$;
$B-V=0^m.759; U-B=0^m.336$) and the check star GSC 3723.104 ($V=12^m.386$;
$B-V=0^m.617; U-B=0^m.408$) were photometrically calibrated relative to North Polar Sequence
(NPS).

During the season of 1998--1999 the photoelectric $BVR$ observations
were carried out with 70-cm reflector of the Sternberg Institute in Moscow
with one-channel photometer constructed by I.M. Volkov and
S.Yu. Shu\-ga\-rov. The photomultiplier FEU-79 was used to repeat the
response close to that of Johnson's $R_J$ band. The $BVR$ magnitudes
of the same reference stars were measured relative to stars of %supporting%
catalogue [21]. The $B$ and $V$ magnitudes of reference stars were found
to be close to previously determined ones within the limits of few hundredth
of magnitude, and  the $R$ magnitudes were $9^m.884$ and $12^m.034$, accordingly for
comparison and check stars. The characteristic accuracy of a single
measurement was of $0^m.02-0^m.03$.

In 2001 January we took a set of CCD observations in $BVR$ bands
with  K585 CCD (530x590 pxl) made by scientific and industrial group
'Electron', using 1-m Zeiss reflector of SAO RAN. The comparison
star GSC 3723.104 situated within the field of the frames was used.
The preliminary reduction were
carried out in the MIDAS environment taking into account the frames of bias,
dark and flat. Our photometry was performed with a V.P. Goranskij's
software based on the method of corrected aperture measurements and
operating in OS Windows NT. The results agree well with the photoelectric
photometry, but have a higher accuracy of $0^m.005$ in $V$ and $R$ bands.
We have carried out monitoring in $B$ band continued for $27^m$ and $28^m$
in two nights on JD 2451928 and 2451929 with 40$^s$ exposures.
The dispersion of measurements was $0^m.008$ and $0^m.007$ for these nights,
accordingly. Such a small dispersion confirmed the complete absence
of rapid variability within the scale of minutes and tens of  minutes
or flickering characteristic of some symbiotic variables.

Additionally, the published data in $UBVR$ bands from [16,19,22,23]
were used for our analysis. The systematic differences between the individual
observing sets of $UBV$ bands were determined using simultaneous or
close by time observations, and these differences  were taken into
consideration in the way to reduce all the data to a single homogeneous
Crimean set obtained by N.V. Metlova. The results of our observations
are collected in Table 1, and the light curves in quiescence are shown in Fig.~1.

The cycle of sinusoidal light variation is seen in Fig.~1 in all the bands
with the amplitudes of $0^m.2U,\ 0^m.15B$ and $0^m.12V$, with a broad
brightness maximum around JD 2451270, and the more narrow minimum around JD 2451820.
The duration of the cycle is $1100^d\pm50^d$. The dispersion of
the $V$ light curve is very small, it is only of $0^m.029$. The behaviour of
the star after the outburst differs radically from that one observed before the
outburst as appears from observations [6], although at an average the levels
of brightness have not changed in quiescence. Our data show  a clear correlation
of brightness changes in all the bands. For instance, the correlation
coefficient between $B$ and $V$ is equal to 0.84, and that between
$U$ and $B$ is equal to 0.82. These coefficients are equal only to 0.58
and 0.37, correspondingly, for observations in [6]. Probably, the observations
in [6] have a lower accuracy. Such a cyclic behaviour of the star
after its outburst resembles that of some symbiotic stars revealing
the periodic variability due to the reflection effect. In such cases, the
period of variability coincides usually with the orbital period.

To verify the $11^d.7$ day period found by Miroshnichenko [5]
in the observations from [6] with our photometry, at first, we substracted
the mean smoothed light curve of the $1100^d$ cycle from all $UBV$ band
light curves. This procedure called prewhitening is widely used
in the frequency analysis. Then the residuals were subjected to
frequency analysis, making use of Deeming [24] method.
The period by Miroshnichenko may be confirmed
only by our $V$ band observations. The dispersion of the residuals is only
$0^m.029$ in this band. The amplitude spectrum  of the harmonic
component is shown in Fig.~2.
The estimates of significance levels for the peaks in the spectrum
are revealed from the analysis of the amplitude
distribution function of random series calculated from the initial
series of residuals  by shuffling among its magnitudes. The amplitude
of the peak concerning to the 11-day period is equal only to $0^m.015$.
The light curve is shown in Fig.~3. The significance level of this
period is very high and exceeds the value of 99.999\%, inspite of the fact
that the amplitude of our light curve ($0^m.03$) is at least 5 times
less then that one in [5]. The dispersion of our light curve is
$0^m.024$, so the conribution of the periodic component to the residuals is only
$\approx$20 per cent. The phases of the light curve given in Fig.~3
are calculated with the formula:

$Max = 2451075.56 + 11^d.719^.E$\ \ \ \ \ (1).

The value of period has been defined with the accuracy of $\pm0^d.02$.
Undoubtedly the period is real due to coincidence with the value
found by Miroshnichenko with the independent set of data,
although no significant peaks have been found in the frequency spectra
of $B$ and $U$ band data. It is assumed that the observations in
$V$ band are more accurate than in other bands. Using again the
photometry published in [6], we have revealed the following formula valid
before the outburst:

$Max = 2448995.64 + 11^d.704^.E$\ \ \ \ \  (2).

Here, the period value may be defined in the range from $11^d.69$ to $11^d.72$.
Then the amplitude of the periodic component in the light curve
was  $0^m.16~B$, $0^m.13~V$ and $0^m.10~R$. The shape of the light
curve in $B$ and $R$ bands was close to a sinusoid. It is
two-humped in $V$ band (see Fig.~4 in [5]). Phases calculated with
formulae (1) and (2) do not agree, but the agreement of light curves before
the outburst and after it may
be achieved with two alternative period values $11^d.69$ and $11^d.75$.
It should be noted, that in [25] the $11^d$.7 period was not confirmed
with CCD photometry in 1999.

We have studied the possibility of connection between the photometric
11-day period and the period of jet rotation observed by methods of
radio interferometry. The VLA radio map of CI Cam was published in [17].
The date of this map is not given, but its dimensions are 0.6x0.6
arcseconds. Two opposite S-shaped jets are seen in the map,
the size of each one reaches $0''.3$,  measuring from the central source.
The structure corresponds approximately to a single revolution of
corkscrewing spiral located in a conic surface. If we adopt the proper motion
of each jet of 0.026 arcseconds per day from [18], we will get
the time of a single revolution of $11^d.5\pm1^d.0$, which is very
close to the value of period by Miroshnichenko. If connection between
radio and optical periods exists, the 11-day period may be treated as
the rotation period of the star with a dipole magnetic field,
the axis of the magnetic dipole inclining to the axis of rotation,
and with  jets to be ejected just along the axis of the magnetic dipole.

The model of rotating jets was constructed to verify this hypothesis.
A computer software created earlier to calculate the structure of SS~433 jet
was used. Four parameters were searched to describe the shape of jets seen
in the radio map [17]: $i$ is the inclination angle between the  axis of star
rotation and the line of sight, $\theta$ is the angle between the rotation axis
and the direction of the jet outflow, $\psi$ is the rotation phase of the
moment when the jet outflow had begun, and the fourth parameter is the date
of the radio map.
$\psi$ parameter is an analogue of the precession phase for the jet of SS~433.
It is  counted out from the moment when the angle between the jet and the line
of sight was minimum. According to our hypothesis, just this moment
the brightness of the rotating star had to be the highest one. The rotation period
was assumed to be equal to $11^d.7$.

The narrow ranges of the angles $i = (35-40^o)$ and $\theta = (7-10^o)$
were found as a result of the modelling. Using these angles, we have
calculated the real velocity of jets in space which is equal to 0.23--0.26$c$,
if the projection of this velocity on the plane of the sky is 0.15$c$ [18].
One may consider 1998 March 31.6~UT (JD 2450904.1), when the X-ray outburst
has started [9], as the start of  jet  outflow. The calculated
rotation phase $\psi$ of the outflow start depends on what the fraction of jet
revolution occured by the moment of taking the radio map, and therefore
on the date of this map. The radio map may be reproduced by our model
already at 78 per cent of the jet revolution, that corresponds to
a minimum date of 1998 April 9.4~UT and to $\psi = 0.69$. A complete
revolution would correspond to date of 1998 April 12.6~UT
and to $\psi = 0.74$. The inclination angle $i$ may be determined up to
its sign, and the $\psi$ phase may have  additional
values in the range of the $\psi = 0.19 - 0.24$ due to the uncertainty of choice
of the jet directed toward the observer. The calculated $\psi$ phases of the
X-ray outburst start are 0.37 from formula (1), and 0.06 from
formula (2), and both do not agree with the calculated ranges of the $\psi$
parameter.

The main conclusion of our  modelling of jets is that the radio map [18]
may be represented by the model of jets rotating with the 11.7-day period
using a specific set of parameters. Moreover, the modelling may
relate the appearence of the jet with the phase of the 11-day
period light curve. This will become possible after publication of  VLA radio maps
with the specific dates of observations, and after a  correction of 11-day
period.

\section*{Spectral observations}

Spectral observations of CI Cam began with the 6-m BTA telescope
at the outburst on 1998 April 4 using the SP-124 spectrograph,
and continued till 2001 January 27 mainly with 1-m Zeiss reflector
of SAO and the UAGS spectrograph.
The spectra were registered with different CCDs. The primary reduction was
made in MIDAS environment taking into account the frames of bias and
darkness. All the spectra taken during individual nights were put
together.
They were calibrated to wavelength scale with the reference spectra of
Ne-Ar lamp. The sky background was exposed through the same slit as a
star from opposite sides of star spectrum, it was  summed up by frame rows, and
then subtracted from star spectrum. A series of spectra on 2000 December 6
was not taken with a slit, but taken with an aperture, and the night sky
spectrum was not exposed separately. These spectra were used only
to measure the radial velocity. Finally, all  spectra were normalized
to continuum, the value of which was accepted as a unit. All spectral
reductions were performed  by E.A.~Barsukova. A total list of spectra along with
their dates, spectral range and resolution, number of frames per night,
lists of telescopes and of observers are given in Table~2.

The results of our spectral observations in the outburst were
published partly in [13,15].

A typical spectrum in quiescence is shown in Fig.~4a, 4b. It repeats
in detail the spectrum before the outburst published by Downes [4],
and that one after the outburst published by Orlandini et al. [19].
The line identification is given in the figure. We used literature sources
of spectroscopic data and laboratory wavelengths of emission lines
from [26--29]. The bright Balmer emissions, the emissions of neutral
helium, the numerous permitted iron lines, as well as a weak silicon
line at $\lambda$ 6347\AA, and a forbidden emission of nitrogen are seen
in this spectrum. Without any doubt, we identify the emission at
$\lambda$5755\AA\ as a forbidden line of nitrogen [N~II] as distinct from
the identification with Fe~II in [4] and [19]. So, the difference between
the wavelength of this line and laboratory one by 8\AA\ noted in [19] is due to
misidentification. We did not find any enough bright forbidden
line of [Fe~II] in our spectral range; in particular, we identified
the emission at $\lambda \approx$ 4416\AA\ as Fe~II $\lambda$ 4414.78\AA\
as distinct from [Fe~II] in [19]. Thus, because of the absence of [Fe~II]
lines in the optical spectrum of CI~Cam, the star does not correspond
with  the canonic description of B[e] star class in one of its aspects.

At first sight, it is difficult to notice the difference between the spectra
in quescence and ones in outburst (see e.g. [13, 15]). The set of lines
remained the same, they only changed their relative brightness. The most
noticeable spectral changes in the outburst are the following.
The wide pedestals appeared during the outburst in Balmer and He~I lines
which are evidence of the motions of matter with  velocities up to
1200 km/s, although  narrow components of profiles 200--400 km/s
wide seen in quescence remained, too. A bright emission of He~II
$\lambda$4686\AA\ emerged. This emission is very weak  but measurable
in quiescence. The bright emission of Na~I D$_1$ and D$_2$ doublet appeared
as well, it merging with wide He~I $\lambda$ 5876\AA\ line, and looking like
a hump in its profile. As the brightness declined, and width of He~I line
decreased, the sodium emission got detached from the helium line, and vanished then.
It has a complicated profile in the high resolution spectrum, containing
emission and absorption components, clearly the last ones have
interstellar origin. On the contrary, the forbidden line of
[N~II] $\lambda$ 5755\AA\, almost invisible during the outburst, became one of
the brightest lines after the outburst. Note that this line was
seen before the outburst ([4], marked as Fe~II), having the same brightness
as after the outburst.

The contribution of the emission lines to the broad $B$ and $V$ bands (of
the $UBVR_J$ photometric system) is $\approx$10 per cent in quiescence and
$\approx$40 per cent in outburst, but it is respectively of 38 and 58 per cent
to $R_J$ band, which the very bright emission of
H$_\alpha$ is located in.\\

{\it \bf (a) Equivalent widths and fluxes of emission lines}\\

The equivalent widths of bright emission lines marked with filled
squares in Fig.~4a, 4b were measured, and then were re-counted into fluxes
in physical quantities of $erg/cm^2/s$ with the help of the photometric data.
Mainly the brightest lines were chosen, not the blends. In some cases
when a line proved to be a blend and could not be resolved, the total equivalent
width of components was measured. Additionally the ratios of outburst-to-quiescence
values were determined for both equivalent widths, and fluxes.
Mean values of first two nights of observations on April 4 and 5
were adopted as a characteristics of outburst, because the observations
had not been carried in the peak of outburst. Mean values for the
following observing season of 1999 were adopted as a characteristics of
quiescence. The results of observations as  ratios of equivalent
widths,  logarithms of flux ratios, as well as  logarithms of total
excitation potentials (sum of ionization potential and excitation
potential, from [27]) of each line are collected in Table~3.

The typical relations of equivalent widths versus time are shown
in Fig.~5. There, the observations by Orlandini et al. [19] are marked
with  open circles, our ones are marked with filled circles.
The equivalent width measurements from [19] agree well with ours,
what cannot be said about line fluxes. Probably, the considerable
systematic difference in photometry is a cause of disagreement.
The typical relations of line fluxes versus time are shown in Fig.~6.
The run of parameters in the course of the outburst in 1998  is not shown
in Fig.~5 and 6, for better imaging of the quiescent behaviour. Only the
final stage of the outburst decay is seen in the most of plots.
The scales of line variations in outburst may be estimated with the data of
Table~3. Various lines show essentially distinct behaviour. One may
distinguish at least 5 types of spectral line behaviour, these types
are given in the last column of Table~3.

{\it The He~II type}

A strong outburst of the emission with equivalent width variation at least
of 133 times, and flux variation of 300 times relative to quiescence occured.
It is not shown in the figure due to inconvenience of presentation.
Probably, the He~II line originated  in immediate vicinity of
the compact object.

{\it The H$\alpha$ type}

The line flux outburst by 5--10 times was observed, then sinusoidal change of equivalent
width reproduced the cycle of brightness variability in detail (Fig.~5a).
The correlation  between equivalent widths of lines of this type
and brightness in $UBVR$ bands exists. Balmer lines, and a majority of
Fe~II lines vary in such a way. The equivalent width variations of these lines
do independently confirm
the photometric 1100-day cycle. As a result of conversion of equivalent
widths to fluxes, the relative amplitude of cyclic variation increases from
25 to 43 per cent (see Fig.~6, H$_\beta$).
Undoubtedly, the part of Balmer emission belongs to gaseous envelope of the system.
Probably, once more region of this type line formation exists on the part of
the normal
star surfice faced to the compact companion, nearby the interior Lagrangian point L$_1$.
It is just the region which may be responsible for the line brightness variations with
the phase of the orbital period.

{\it The He~I type}

The line flux outburst by 15--50 times took place, then strong irregular changes of equivalent
widths were detected in quiescence  (Fig.~5b). Besides all He~I lines, two lines
of the Fe~II $\lambda$ 6318 and 6385\AA\ belong to this type.
Cyclic 1100-day variation was not observed.
The lines of He~I type showed gradual systematic weakening of the flux and the equivalent
widths in the season of 1999 (JD 2451399--2451485), when the H$_\alpha$ type lines
had a wide maximum of the dependence of equivalent widths upon time. Note that the X-ray source
was detected with BeppoSAX sattelite having a soft spectrum in the
beginning of this season, and later having a hard spectrum in the middle
of the season [19, 20]. Probably, the behaviour of such type lines is
sensitive to the X-ray flux from the compact companion, and we have observed the
decay of line fluxes
after a weak local X-ray flare. So,  the He~I type lines may be
formed close to the compact companion, possibly, in the accretion disk.

{\it The Si~II type}

The line intensity outburst by 6-18 times was followed by a gradual slow decay of
the brightness. No trace of cyclic 1100-day variation is apparently observed. Besides
the Si~II $\lambda$6347\AA\ two line of  Fe~II $\lambda$6148 and 6492\AA\
vary in such a way (see Fig.~5c, three lower curves).

{\it The [N~II] type}

The [N~II] $\lambda$5755\AA\ line finds itself both in the blue and in the red
spectra, therefore its variation has been traced in more detail
(Fig.~5c, the upper curve). The emission was very weak but measurable
during the outburst, and its equivalent width was gradually increasing
in the course of star brightness decay (Fig.~5c, the upper curve).
It reached maximum in the end of the outburst,
and then decreased gradually. Nevertheless, the flux in this line
was constant in the outburst during 50 days, but in 250 days after the X-ray peak
it increased by one and a half times. That time all other line fluxes
had  already decreased to their quiescent levels. Later, by the beginning of 2001
the flux of the [N~II] line fell down to its outburst level.

This forbidden line usually forms  in the
most rarefied medium, which can exist only in the exterior edge of so
dense gas and dust envelope as that observed for CI Cam. If we assume that
the line brightening is caused by the outburst radiation wave arriving at
outlying parts of the envelope, than the
exterior envelope radius shoud be equal to {\it at least} 50 light days,
what is of $5-8''$ for the known distance to the star. No nebulosity is
observed on the deep limiting frames taken with broad band filters.
The profiles of brightness
distribution measured accurately along the slit of our spectra from 6-m telescope
with angular resolutions of 2--3 arcseconds both in the [N~II] $\lambda$5755\AA\
emission and in the continuum do not differ absolutely, the FWHM coinciding
with the accuracy of 3--5 per cent! So, the above assumption can not be
accepted.

An alternative assumption is that the emission's brightening in  50 --
250 days after the outburst is caused by ejection of rarefied matter beyond the
envelope range when some density wave generated by the outburst reaches
to the exterior edge of the envelope. The velocity of wave motion
may be derived as 1200 km/s using pedestal widths of emission lines
during the outburst [15, 23]. Then the outer radius of envelope should be
at least $0''.02-0''.03$ what is beyond the limit of accuracy of our
along slit spectra measurements . Gaseous medium with the density of
$10^5-10^6$ atoms/cm$^3$ may  extend up to this distance.

The existence of five types of spectral line behaviour confirms that
a stratified envelope exists in the system. The $H_\alpha$
type of a cyclic 1100-day variation of equivalent width of some lines
indicates the presence of a giant star in the system, this giant
having the reflection effect on the
region of its surface faced to the compact object. The relation
between the line flux outburst-to-quiescence ratio and the total line excitation
potential is shown in Fig.~8 in logarithmic scale. The clear dependence
between the outburst amplitude and the excitation potential confirms the
envelope stratification; in the envelope
the closer the location of radiation source and the stronger
the source, the higher excited and ionized matter is.\\

{\it \bf (b) Radial velocities}\\

In the course of radial velocity calibration of spectra with the UAGS
spectrograph of the 1-m Zeiss telescope  we have found the sighificant
systematic errors of the
dispersion curves changing from night to night by 2--3\AA. Probably,
they are due to device bending. Nevertheless, the internal accuracy
of dispersion curve approximation is always equal to  0.2--0.4\AA,
or 10--20 km/s in the velocity scale. To minimize systematic errors
of our radial velocities we referred the velocity scale to telluric
lines. The bright sky line of [O~I]\ $\lambda$5577\AA\ was used in
the blue spectra, and the head of O$_2\ \lambda$6872\AA\ atmospheric
absorption band was used in the red ones. This band  does not
split into components at our spectral resolution, seems symmetric enough,
and may be a good bench mark. The systematic difference of
radial velocity level between blue and red parts of spectra may
be caused by uncertainty of the effective wavelength of the O$_2$ band.
Both the line of sky and the absorption band are well seen in
all the spectra. In such a way we have obtained the uniform sets
of radial velocities for  H$_\alpha$, [N~II] $\lambda$ 5755\AA\, and
for several Fe~II and He~I lines, closely spaced in the spectra to
selected reference lines. The BTA spectrograms were referred in the
same way to reach a homogeneity. The radial velocity curves of
selected lines are shown in Fig.~7.

H$_\alpha$ line and Fe~II $\lambda$5535 and 6516\AA\ lines do not show any
changes of radial velocity, including the time of outburst. It is clear
if the source of line radiation is located close to inner Lagrangian
point, i.e. close to the mass center of the system. It should be noted that
amplitudes of component velocities are expected to be small in the system having
so long orbital period, as 1100 days. The small radial velocity
variations from season to season are visible in the He~I $\lambda$6678
and 7065\AA\ lines, and a systematic velocity directed to the observer
of 50--100 km/s is noticeable during the outburst. This systematic velocity in
outburst may be due to partial screening of the red part of the line profiles
by matter in the  disk plane inclined to the line of sight. The same
but oppositely directed systematic velocity is seen in
the Fe~II $\lambda$5362\AA\ line at the time of the outburst.
The forbidden [N~II] $\lambda$5755\AA\ line also has a positive velocity deviation
during the season of the most brightness. Probably, these lines are forming
far from the accretion disk, and therefore they behave otherwise than He~I
lines. Our observations give hope that the orbital motion in CI Cam
may by detected in some lines (He~I, for instance) using moderate
resolution spectra. Additional observations at least of one more
possible orbital cycle are needed.

\section*{Summary and conclusions}

On the grounds of photometry and spectroscopy, some details of conception
appear that CI Cam is a stellar system with the orbital period of
1100 days consisting of a giant star and a compact object. The high
temperature continuum may belong to an accretion disk or to  matter
surrounding the compact object. The system is imbedded in a dence
and strongly stratified gas and dust envelope, and resembles
symbiotic systems.  A part of the giant star surface is heated in the
region of the inner Lagrangian point, and the reflection effect is in action
(at least now). Probably, the giant is a late spectral type one,
G8II--K0II [5, 7].

The compact object may be a star having a strong magnetic field, the dipole
axis of which is inclined to the rotation axis, and the rotation period may be
of 11.72 days. In quiescence, an accretion on the compact object from the dence
circumstellar medium goes along the lines of the magnetic field,
so we observe  hot spots or accretive columns in the poles of the dipole like those in polars,
dwarf stellar systems. The contribution of the hot spots to the common light is
small, and the variability amplitude due to the compact object rotation
is small, only about 3 per cent in $V$, and less in other bands.
But it might be 5 times larger before the outburst according to
Miroshnichenko [5]. In outburst ionized matter modifies the magnetic field
structure. The plasma accelerates along the curve dipole lines
up to relativistic velocities ($\approx0.25c$). Coriolis forces
exert considerable pressure upon the lines of the magnetic field, and as a result
the magnetic field lines are rolled up into a collimated tube.
This phenomenon may be a jet collimation mechanism. The syncrotron
radio emission of jets is evidence  that the jet's plasma moves in a
strong magnetic field.

Of course, new additional observations and theoretical calculations are needed
to confirm and specify this conception. So, accurate photometry
and medium-resolution spectroscopy during at least one more 1100-day cycle
are able to show if this cycle of variability reflects the real orbital period.
An analysis of original date-fixed VLA radio maps and of accurate
photometry may give a more precise picture of jet outflow and
clarify a mechanism of an ejection for collimated jets.

The assumption about a white dwarf and a thermonuclear explosion of
matter supplied on its surfice due to accretion has the most grounds
to explain the nature of the compact object and the outburst in April, 1998.
The white dwarf hypothesis is based on the following arguments [19].
First, the two-temperature termal X-ray spectrum in the outburst, and an outburst
duration of about a week being in agreement with Iben's calculation [30]
of a thermonuclear runaway on the surface of a hot white dwarf with
of $\approx1~M_\odot$ mass. [At the same time no X-ray flux is usually
observed in analogous situation in the Novae outbursts --
{\it the authors}]. Second, there is no  rapid variability (in X-rays
and in the optical range). Third, an expanding envelope  in the radio range
is observed that can be explained by ejection of H- and He-rich layers
as a result of the thermonuclear explosion (this is confirmed well
by appearance of the H and He~I lines pedestals in the outburst).
According to [19] the hypothesis of a neutron star is not excluded completely.

The known X-ray novae with a black hole companion exhibit the excess
of power spectrum in the hard X-ray range  and rapid variability of
the X-ray flux, which are not observed in CI Cam. Therefore, the probability
of black hole presence in CI Cam system is very low. The presence
of a normal OB-type star in the system is eliminated in [19], too, because
the expected X-ray flux in quiescence is lower by two orders than observed one
of CI Cam.

The main conclusions of our study are the following.

(1) The 1100-day cycle of variation in CI Cam was found in photometry
as well as in fluxes of some low-excitation emission lines. This cycle
may be a consequence of orbital motion in a widely detached system
with a giant star and of the reflection effect on its surface.

(2) With our $V$-band photometry we confirm the $11^d.7$ period
found by Miroshnichenko [5] before the 1998 outburst. The model calculations
show that this period may be identified with the rotation period of
relativistic jets in radio maps. The radio map modelling reveals the
following range of parameters: the inclination of the jet's rotation axis
to the line of sight $i = (35-40^o)$, the angle between the rotation
axis and the direction of the jet outflow  $\theta = (7-10^o)$, and
the spatial velocity of the jet of 0.23--0.26$c$.

(3) The connection between the behaviour of different emission lines and
their formation regions is found. Apparently this relation  reflects
the star envelope stratification.
The rapid variability of emission fluxes of helium and
several iron lines is found. Probably, the variability is caused by
variations of the X-ray flux in quiescence.
The time lag by 50-250 days of the [N~II] $\lambda$5755\AA\
forbidden line brightening was detected relative to the peak of the outburst
in optical, X-ray and radio ranges.

(4) The amplitudes of the outburst in different emission lines were
studied. The relation  between the  amplitude of the flux
change in lines  and the total line excitation potential was found. \\[3mm]

We are grateful to S.A.~Pustilnik and A.V.~Ugryumov for carrying out
observations of CI Cam in the outburst of 1998 during their
BTA observing time, to staff members of SAO RAN A.N.~Burenkov, G.G.~Valyavin,
V.V.~Vla\-syuk, D.N.~Monin, and N.I.~Serafimovich for their spectral
observations of CI Cam. Also, we are thankful to E.A.~Karitskaya
and S.Yu. Shugarov for their help in photometric observations.
The authors thank P.~Roche and S.Yu.~Shugarov for their opportune information
about the CI Cam outburst.

The study has been supported partially by the
Federal Scientific and Technical Program "Astronomy" through  grant 1.4.2.2.
The authors V.P.~Goranskij and N.V.~Metlova are thankful to this fund
for a  support.

\newpage
\Large{
\centerline{\psfig{figure=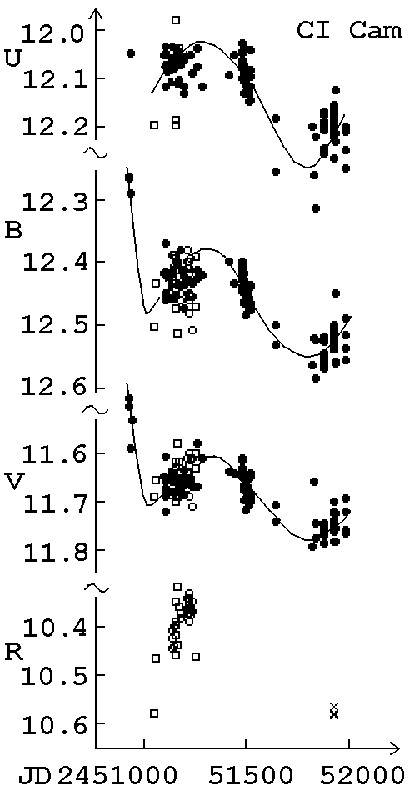,height=18cm,width=10cm}}
Fig.~1. The light curves of CI Cam in $UBVR$ bands\\
\hspace*{4cm} after the outburst in 1998.

\newpage
\centerline{\psfig{figure=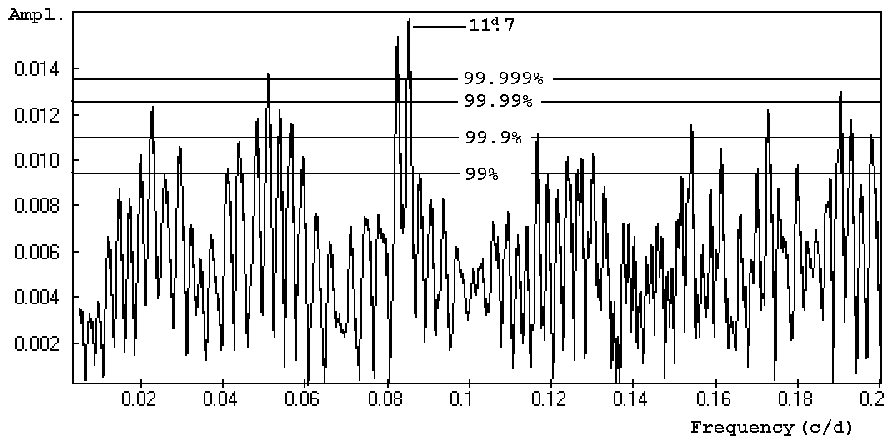,height=8cm,width=15.2cm}}
\begin{center}
Fig.~2. The amplitude -- frequency relation (the amplitude
spectrum according to Deeming) for the residuals
from the mean light curve with 1100-day period in the $V$ band.
The horizontal lines show the significance levels.
\end{center}

\vspace{2cm}
\centerline{\psfig{figure=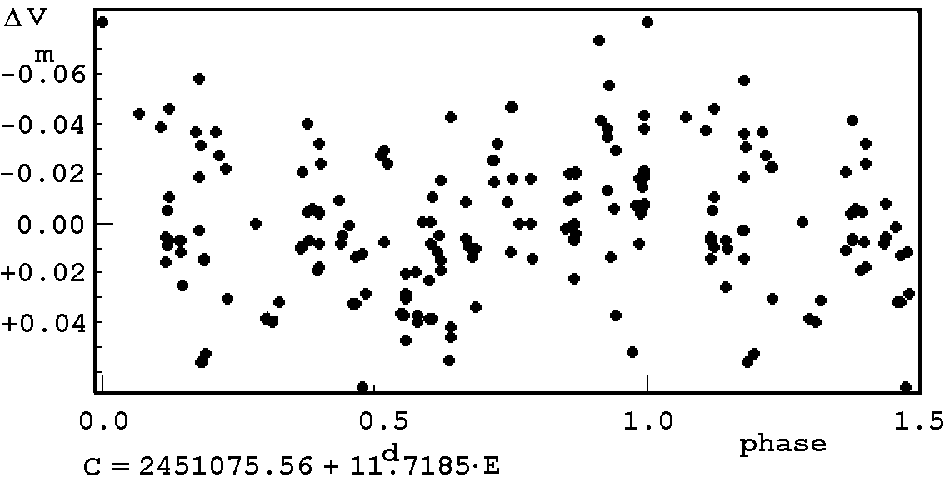,height=8cm,width=15.2cm}}
\begin{center}
Fig.~3. The $V$ band light curve for the residuals with period
11.72 day.
\end{center}

\newpage

\centerline{\psfig{figure=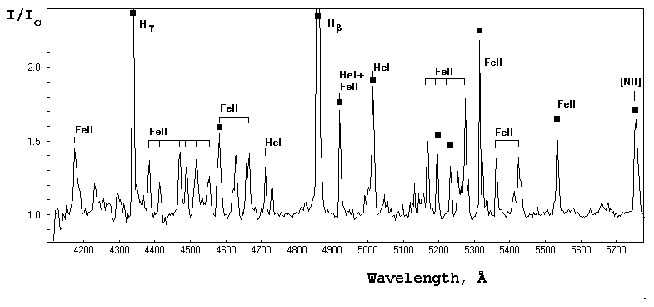,height=8cm,width=15.2cm}}

\centerline{\psfig{figure=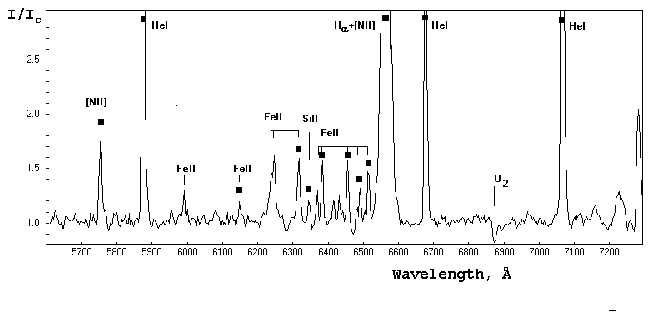,height=8cm,width=15.2cm}}
\begin{center}
Fig.~4. The quiescent spectrum of CI Cam in the blue (a) and red (b)
ranges. The filled squares mark emission lines selected for measurements.
\end{center}

\newpage

\hbox{\hspace{-0.5cm}\psfig{figure=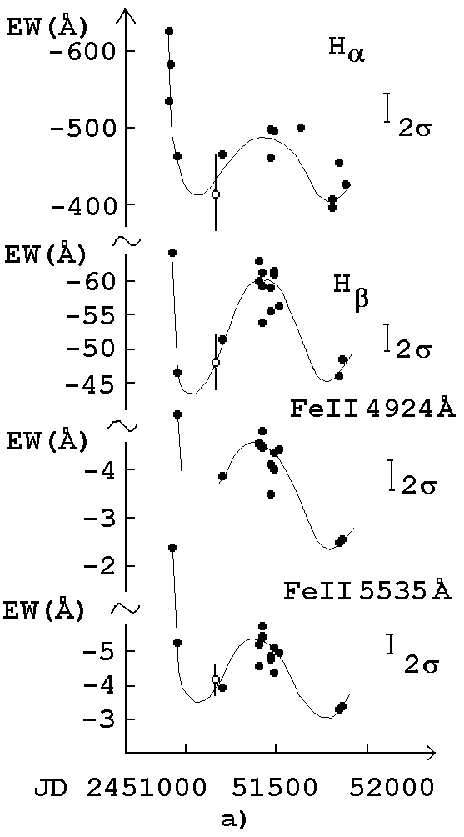,height=9cm,width=5cm}
\hspace{1cm}
\psfig{figure=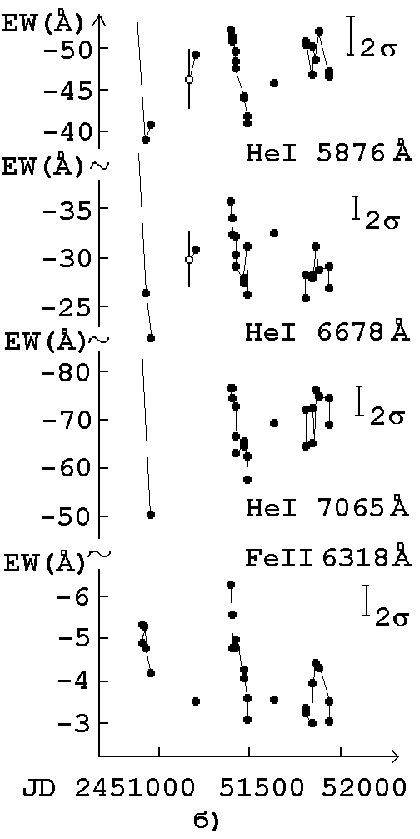,height=9cm,width=5cm}
\hspace{1cm}
\psfig{figure=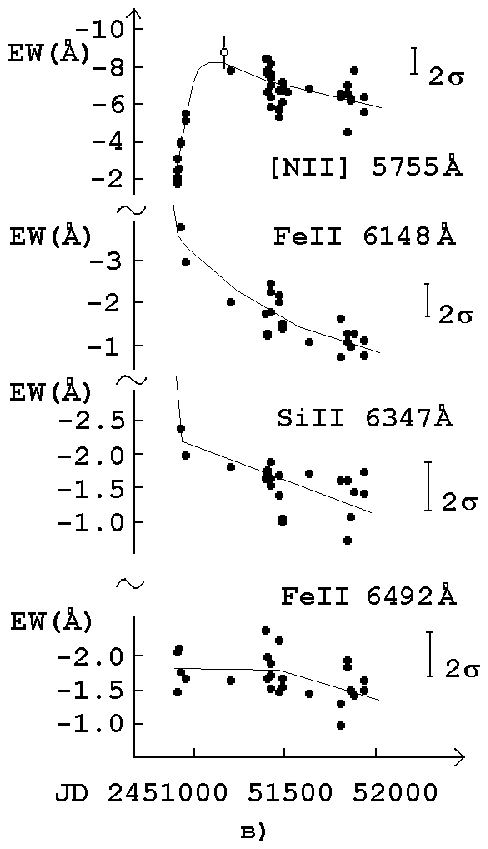,height=9cm,width=5cm}
}
Fig.~5. The relation of equivalent width versus time for some
emission lines. Different types of behaviour are shown: (a)--the H$_\alpha$ type;
(b)--the He~I type; (c, upper curve)--the [N~II] type; (c, three
lower curves)--the Si~II type. The open circles are observations from [19].
The uncertainties of equivalent width determination of 2$\sigma$ are
shown with vertical bars.

\newpage

\centerline{\psfig{figure=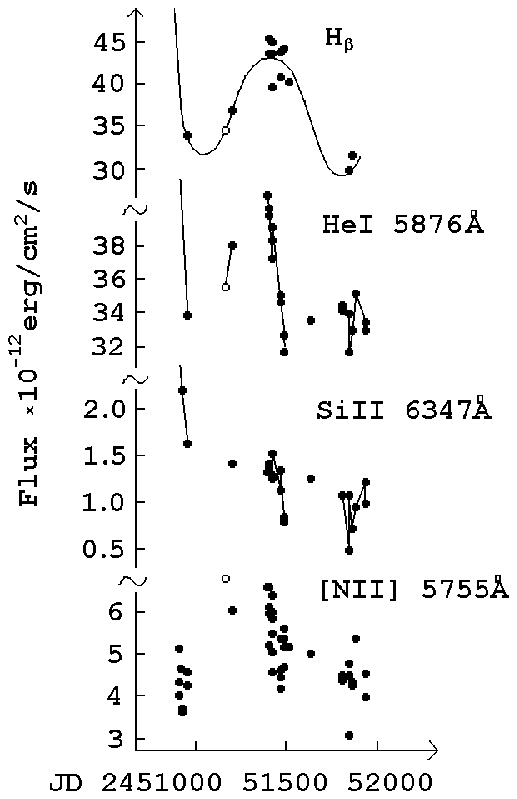,height=18cm,width=13cm}}
\begin{center}
Fig.~6. The relation of fluxes in lines with distinct behaviour types
versus time.
\end{center}

\newpage

\hbox{\hspace{-0.5cm}\psfig{figure=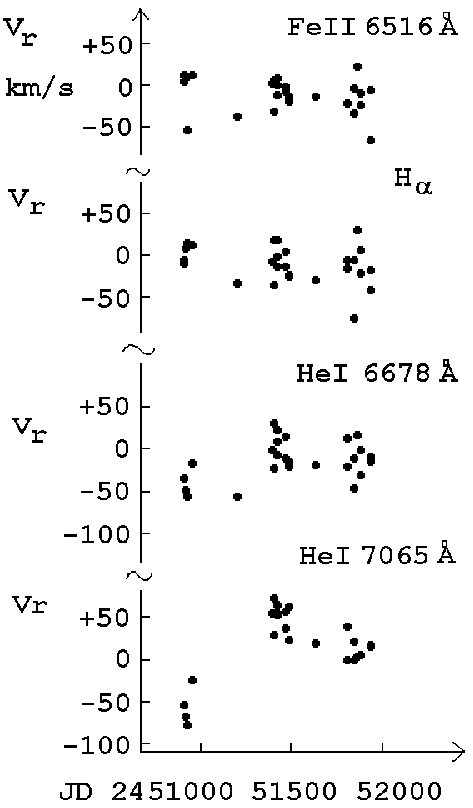,height=13.5cm,width=9cm}
\hspace{1cm}
\psfig{figure=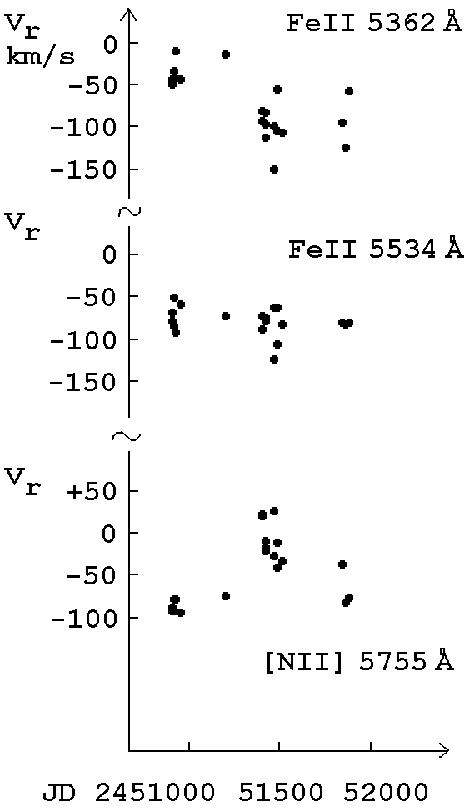,height=13.5cm,width=7.5cm}
}
\begin{center}
Fig.~7. The radial velocity curves of some emission lines:
(a)-- in the red spectral range referred to the
O$_2\ \lambda$6872\AA\ absorption band, (b)-- in the blue  range
referred to the [O~I]\ $\lambda$5577\AA\ telluric line.
\end{center}

\newpage

\centerline{\psfig{figure=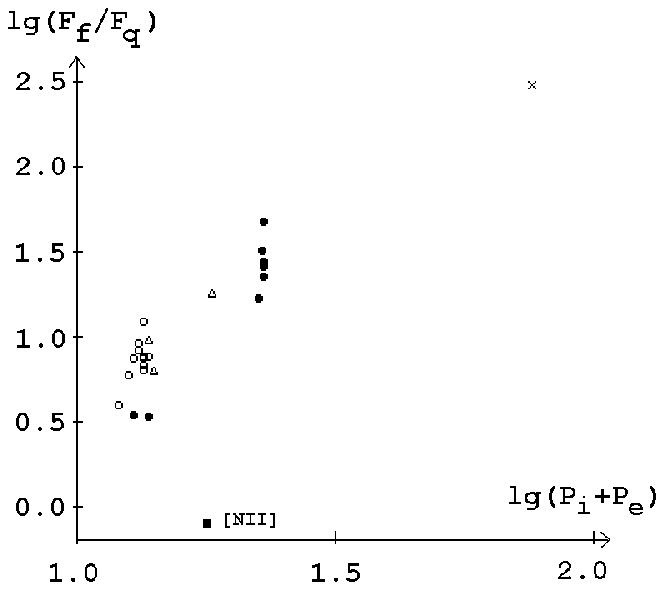,height=13cm,width=15cm}}
Fig.~8. The dependence between the ratio of the emission line outburst
flux ($F_f$) to its quiescent flux ($F_q$) and the line's total excitation potential
(the sum of ionization $P_i$ and excitation $P_e$ potentials), in the
logarithmic scale. The He~II $\lambda$4686\AA\ line is marked with
a cross, He~I and Fe~II lines of the same behaviour type  marked with
filled circles, Balmer lines and Fe~II lines of H$_\alpha$-type are marked
with open circles, the triangles are the Si~II-type behaviour lines,
and the filled square is the [N~II] $\lambda$5755\AA\ line.

}

\newpage
\large{
\begin{center}
Table 1. Photometric observations of CI Cam.\\
\vspace{0.3cm}
\begin{tabular}{cccccccl}
JD hel.   &   V   &   B   &   U   &    R   &Device&Tele-&Obser- \\
24...     &       &       &       &        &      &scope&ver(s) \\
\hline
50907.2757&  9.53 & 10.67~~& -     &   7.89 &CCD&60 &L,G \\
50908.2659& 10.095& 10.967& 10.607&   -    &pe &60 &L,G,M \\
50909.2812& 10.21~~& 11.38~~& -     &   8.66 &CCD&60 &L,G \\
50910.2565& 10.38~~& 11.33~~& -     &   8.84 &CCD&60 &L,G \\
50911.2581& 10.60~~& 11.57~~& -     &   9.11 &CCD&60 &L,G \\
50912.2555& 10.73~~& 11.68~~& -     &   9.25 &CCD&60 &L,G \\
50917.2980& 11.21~~&  -    & -     &   9.86 &CCD&60 &L \\
50918.3140& 11.30~~&  -    & -     &   9.93 &CCD&60 &L \\
50920.2840& 11.344& 12.141& 11.741&      - &pe &60 &M \\
50922.2995& 11.385& 12.172& 11.783&      - &pe &60 &M \\
50927.2826& 11.438& 12.213& 11.808&      - &pe &60 &M \\
50929.2743& 11.488& 12.264& 11.908&      - &pe &60 &M \\
50933.2732& 11.504& 12.265& 11.897&      - &pe &60 &M \\
50934.3195&   -   &   -   & 11.921&      - &pe &60 &M \\
50941.2947& 11.59~~& 12.29~~& 12.05~~&      - &pe &60 &M \\
50942.2832& 11.53~~& 12.38~~& 12.16~~&      - &pe &60 &M \\
51103.4740& 11.648& 12.418& 12.051&      - &pe &60 &M \\
51104.5430& 11.720& 12.461& 12.117&      - &pe &60 &M \\
51105.4780& 11.690& 12.450& 12.051&      - &pe &60 &M \\
51105.4850& 11.673& 12.461& 12.077&      - &pe &60 &M \\
51105.4920& 11.683& 12.458& 12.071&      - &pe &60 &M \\
51110.5100& 11.659& 12.414& 12.050&      - &pe &60 &M \\
51111.5070& 11.607& 12.369& 12.036&      - &pe &60 &M \\
51112.3910& 11.675& 12.422& 12.086&      - &pe &60 &M \\
51133.2798& 11.606& 12.410& 12.086&      - &pe &60 &S \\
51133.2960& 11.630& 12.398& 12.062&      - &pe &60 &S \\
51133.3062& 11.589& 12.408& 12.038&      - &pe &60 &S \\
51137.4366& 11.644& 12.417& 12.035&      - &pe &60 &M \\
51141.3537& 11.655& 12.416& 12.066&      - &pe &60 &M \\
51143.3548& 11.626& 12.398&   -   &  10.407&pe &70 &G \\
51143.3705& 11.644& 12.434&   -   &  10.424&pe &70 &G \\
51143.3857& 11.658& 12.468&   -   &  10.443&pe &70 &G \\
51143.3548& 11.626& 12.398&   -   &  10.407&pe &70 &G \\
51143.3705& 11.644& 12.434&   -   &  10.424&pe &70 &G \\
51143.3857& 11.658& 12.468&   -   &  10.443&pe &70 &G \\
51148.3629& 11.607& 12.398&   -   &  10.398&pe &70 &G \\
51148.3791& 11.617& 12.417&   -   &  10.440&pe &70 &G \\
51152.3388& 11.681& 12.457& 12.079&      - &pe &60 &M \\
51161.3109& 11.681& 12.435& 12.110&      - &pe &60 &M \\
51163.5998& 11.660& 12.435& 12.040&      - &pe &60 &M \\
51164.4602& 11.652& 12.399& 12.062&      - &pe &60 &M \\
51176.2600& 11.677& 12.436& 12.103&      - &pe &60 &M \\
51179.3464& 11.633& 12.460&   -   &  10.380&pe &70 &G \\
51179.3608& 11.643& 12.470&   -   &  10.370&pe &70 &G \\
51180.2383& 11.666& 12.379& 12.072&      - &pe &60 &M \\
\hline
\end{tabular}

\newpage
Table 1 (continued). \\
\vspace{0.3cm}
\begin{tabular}{cccccccl}
JD hel.   &   V   &   B   &   U   &    R   &Device&Tele-&Obser-\\
24...     &       &       &       &        &      &scope&ver(s)\\
\hline
51197.3893& 11.649& 12.413& 12.055&      - &pe &60 &M \\
51197.4001& 11.636& 12.437& 12.131&      - &pe &60 &M \\
51199.2277& 11.686& 12.472& 12.116&      - &pe &60 &M \\
51199.2378& 11.667& 12.424& 12.061&      - &pe &60 &M \\
51213.2166& 11.623& 12.390&   -   &      - &pe &70 &G \\
51213.2207& 11.643& 12.410&   -   &  10.360&pe &70 &G\\
51213.2389& 11.613& 12.410&   -   &  10.370&pe &70 &G\\
51213.2490& 11.583& 12.420&   -   &  10.340&pe &70 &G\\
51218.2389& 11.570& 12.459&   -   &  10.362&pe &70 &G,K\\
51218.2464& 11.609& 12.464&   -   &  10.389&pe &70 &G,K\\
51218.2540& 11.631& 12.428&   -   &  10.348&pe &70 &G,K\\
51218.2612& 11.596& 12.463&   -   &  10.354&pe &70 &G,K\\
51223.2283& 11.663& 12.470&   -   &  10.330&pe &70 &G,K\\
51223.2329& 11.663& 12.470&   -   &  10.330&pe &70 &G,K\\
51223.2425& 11.623& 12.410&   -   &  10.340&pe &70 &G,K\\
51223.2529& 11.633& 12.490&   -   &  10.370&pe &70 &G,K\\
51223.2630& 11.613& 12.420&   -   &  10.340&pe &70 &G,K\\
51227.2442& 11.662& 12.431& 12.053&    -   &pe &60 &M\\
51232.2840& 11.585& 12.404&   -   &  10.365&pe &70 &G\\
51232.2974& 11.633& 12.465&   -   &  10.373&pe &70 &G\\
51232.3059& 11.622& 12.441&   -   &  10.358&pe &70 &G\\
51235.3438& 11.668& 12.433& 12.091&    -   &pe &60 &M\\
51235.3636& 11.583& 12.464&   -   &  10.347&pe &70 &G\\
51235.3771& 11.681& 12.517&   -   &  10.369&pe &70 &G\\
51264.3012& 11.579& 12.413& 12.037&      - &pe &60 &M\\
51265.2567& 11.669& 12.426& 12.077&      - &pe &60 &M\\
51287.2855& 11.610& 12.420& 12.117&      - &pe &60 &M\\
51420.5427& 11.640& 12.400& 12.095&      - &pe &60 &M\\
51440.5108& 11.643& 12.433& 12.051&      - &pe &60 &M\\
51485.5330& 11.647& 12.429& 12.061&      - &pe &60 &M\\
51485.5639& 11.638& 12.421& 12.049&      - &pe &60 &M\\
51485.5708& 11.649& 12.429& 12.079&      - &pe &60 &M\\
51485.5771& 11.648& 12.438& 12.066&      - &pe &60 &M\\
51485.5837& 11.634& 12.418& 12.029&      - &pe &60 &M\\
51485.5910& 11.635& 12.406& 12.067&      - &pe &60 &M\\
51485.5972& 11.615& 12.417& 12.101&      - &pe &60 &M\\
51485.6048& 11.634& 12.447& 12.045&      - &pe &60 &M\\
51485.6111& 11.633& 12.400& 12.050&      - &pe &60 &M\\
51485.6180& 11.610& 12.417& 12.071&      - &pe &60 &M\\
51485.6246& 11.634& 12.416& 12.046&      - &pe &60 &M\\
51485.6319& 11.647& 12.441& 12.063&      - &pe &60 &M\\
51492.4350& 11.698& 12.466& 12.107&      - &pe &60 &M\\
51492.4461& 11.678& 12.449& 12.102&      - &pe &60 &M\\
51492.4572& 11.696& 12.463& 12.132&      - &pe &60 &M\\
51493.5305& 11.651& 12.418& 12.072&      - &pe &60 &M\\
\hline
\end{tabular}
\end{center}

\newpage
\begin{center}
Table 1 (continued).\\
\vspace{0.3cm}
\begin{tabular}{cccccccl}
JD hel.   &   V   &   B   &   U   &    R   &Device&Tele-&Obser-\\
24...     &       &       &       &        &      &scope&ver(s)\\
\hline
51493.5371& 11.666& 12.429& 12.039&      - &pe &60 &M\\
51493.6246& 11.674& 12.434& 12.095&      - &pe &60 &M\\
51493.6322& 11.670& 12.449& 12.107&      - &pe &60 &M\\
51499.5344& 11.679& 12.458& 12.085&      - &pe &60 &M\\
51499.5414& 11.719& 12.485& 12.114&      - &pe &60 &M\\
51502.5240& 11.657& 12.449& 12.102&      - &pe &60 &M\\
51502.5313& 11.674& 12.460& 12.097&      - &pe &60 &M\\
51502.5386& 11.671& 12.461& 12.102&      - &pe &60 &M\\
51514.5609& 11.707& 12.472& 12.132&      - &pe &60 &M\\
51514.5679& 11.687& 12.458& 12.150&      - &pe &60 &M\\
51525.5179& 11.691& 12.440& 12.040&      - &pe &60 &M\\
51525.5249& 11.649& 12.436& 12.084&      - &pe &60 &M\\
51525.5315& 11.642& 12.471& 12.097&      - &pe &60 &M\\
51526.4565& 11.702& 12.475& 12.144&      - &pe &60 &M\\
51526.4631& 11.685& 12.465& 12.118&      - &pe &60 &M\\
51645.2979& 11.743& 12.499& 12.185&      - &pe &60 &M\\
51645.3048& 11.706& 12.531& 12.295&      - &pe &60 &M\\
51822.6055& 11.794& 12.563& 12.200&      - &pe &60 &M\\
51831.4368& 11.660& 12.521& 12.301&      - &pe &60 &M\\
51836.4257& 11.746& 12.523& 12.220&      - &pe &60 &M\\
51842.6180& 11.776& 12.585& 12.370&      - &pe &60 &M\\
51842.6246& 11.808& 12.647& 12.506&      - &pe &60 &M\\
51882.5555& 11.755& 12.554& 12.244&      - &pe &60 &M\\
51882.5625& 11.787& 12.557& 12.256&      - &pe &60 &M\\
51882.5694& 11.766& 12.544& 12.170&      - &pe &60 &M\\
51882.5760& 11.765& 12.565& 12.207&      - &pe &60 &M\\
51882.5829& 11.771& 12.568& 12.197&      - &pe &60 &M\\
51882.5895& 11.769& 12.518& 12.191&      - &pe &60 &M\\
51882.5965& 11.744& 12.526& 12.177&      - &pe &60 &M\\
51925.4042& 11.721& 12.501& 12.157&      - &pe &60 &M\\
51925.4118& 11.725& 12.521& 12.180&      - &pe &60 &M\\
51926.5173& 11.764& 12.538& 12.196&      - &pe &60 &M\\
51926.5260& 11.754& 12.534& 12.268&      - &pe &60 &M\\
51927.2860& 11.760& 12.534& 12.188&      - &pe &60 &M\\
51927.2920& 11.782& 12.526& 12.166&      - &pe &60 &M\\
51928.2664& 11.749& 12.557&   -   &  10.566&CCD&100&B,G\\
51928.2674& 11.749& 12.557&   -   &  10.566&CCD&100&B,G\\
51929.3068& 11.751& 12.537&   -   &  10.579&CCD&100&B,G\\
51930.2810& 11.742& 12.557&   -   &  10.582&CCD&100&B,G\\
51932.3709& 11.762& 12.530& 12.187&      - &pe &60 &M\\
51932.3782& 11.748& 12.530& 12.222&      - &pe &60 &M\\
51932.3851& 11.753& 12.530& 12.187&      - &pe &60 &M\\
51932.3914& 11.754& 12.522& 12.191&      - &pe &60 &M\\
51932.3987& 11.748& 12.526& 12.178&      - &pe &60 &M\\
51932.4053& 11.757& 12.532& 12.212&      - &pe &60 &M\\
\hline
\end{tabular}

\newpage
Table 1 (the end). \\
\vspace{0.3cm}
\begin{tabular}{cccccccl}
JD hel.   &   V   &   B   &   U   &    R   &Device&Tele-&Obser-\\
24...     &       &       &       &        &      &scope&ver(s)\\
\hline
51932.4126& 11.742& 12.508& 12.214&      - &pe &60 &M\\
51932.4195& 11.737& 12.520& 12.200&      - &pe &60 &M\\
51932.4264& 11.701& 12.519& 12.187&      - &pe &60 &M\\
51934.5163& 11.785& 12.561& 12.230&      - &pe &60 &M\\
51935.2836& 11.757& 12.559& 12.213&      - &pe &60 &M\\
51935.2898& 11.726& 12.449& 12.126&      - &pe &60 &M\\
\hline
\end{tabular}
\end{center}

\begin{tabular}{lll}
Observers:  \\
B -- Barsukova E.A.         &L -- Lyuty V.M.  \\
G -- Goranskij V.P.         &M -- Metlova N.V. \\
K -- Karitskaya E.A.(Institute of Astronomy, RAN)&S -- Shugarov S.Yu.(SAI)\\
\end{tabular} \\[2mm]

Telescopes:\\
60 -- 60-cm Zeiss reflector of SAI Crimean station;\\
70 -- 70-cm reflector of SAI in Moscow;\\
100-- 100-cm Zeiss reflector of SAO RAN.\\[2mm]

Devices:\\
CCD -- CCDs SBIG ST-6 and Electron-K585 (see text);\\
pe  -- photoelectric one-channell UBV- and BVR- photometers (see text).
\newpage
\centerline{ Table 2. Spectral observations.}

\begin{center}
\begin{tabular}{lcccrcl}
JD hel.  &    Date   &   Range   &Resolu-  &Number&Tele- &Observers    \\
24...    &           &   (\AA)   &tion(\AA)&of sp.&scope &             \\
\hline
50908.259& 1998.04.04& 3800-6100 &    4    &  2   & 6-m  & P,U       \\
50908.260& 1998.04.04& 5000-7400 &    4    &  1   & 6-m  & P,U       \\
50909.176& 1998.04.05& 5000-7400 &    4    &  1   & 6-m  & P,U       \\
50909.20 & 1998.04.05& 3800-6100 &    4    &  1   & 6-m  & P,U       \\
50910.20 & 1998.04.06& 3800-6100 &    4    &  2   & 6-m  & P,U       \\
50910.205& 1998.04.06& 5000-7400 &    4    &  3   & 6-m  & P,U       \\
50923.25 & 1998.04.19& 3800-6100 &    7    & 29   & 6-m  & P,U,M    \\
50923.266& 1998.04.19& 5000-7400 &    7    &  2   & 6-m  & P,U,M    \\
50950.30 & 1998.05.16& 3800-6100 &    4    &  2   & 6-m  & V1,F       \\
50950.314& 1998.05.16& 5000-7400 &    4    &  2   & 6-m  & V1,F       \\
51204.333& 1999.01.25& 3800-6100 &    4    &100   & 6-m  & B1,V1,M,F \\
51206.30 & 1999.01.27& 5000-7700 &    0.23 &  3   & 6-m  & B1,M       \\
51398.534& 1999.01.08& 5580-7320 &    4    &  2   & 1-m  & B2          \\
51399.538& 1999.08.09& 4100-5800 &    4    &  2   & 1-m  & B2          \\
51399.525& 1999.08.09& 5580-7290 &    4    &  2   & 1-m  & B2          \\
51400.510& 1999.08.10& 4100-5800 &    4    &  4   & 1-m  & B2          \\
51400.531& 1999.08.10& 5600-7300 &    4    &  4   & 1-m  & B2          \\
51423.551& 1999.09.02& 5590-7330 &    4    &  4   & 1-m  & B2          \\
51423.566& 1999.09.02& 4100-5780 &    4    &  3   & 1-m  & B2          \\
51424.551& 1999.09.03& 4100-5780 &    4    &  4   & 1-m  & B1,B2       \\
51425.581& 1999.09.04& 5600-7300 &    4    & 13   & 1-m  & B1,B2       \\
51426.519& 1999.09.05& 4100-5780 &    4    &  3   & 1-m  & B1,B2       \\
51426.561& 1999.09.05& 5600-7300 &    4    &  5   & 1-m  & B1,B2       \\
51464.523& 1999.10.13& 5600-7320 &    4    &  6   & 1-m  & B3         \\
51464.542& 1999.10.13& 4100-5780 &    4    &  4   & 1-m  & B3         \\
51465.434& 1999.10.14& 4100-5780 &    4    &  4   & 1-m  & B3         \\
51465.458& 1999.10.14& 5600-7320 &    4    &  6   & 1-m  & B3         \\
51485.653& 1999.11.03& 4100-5700 &    4    &  3   & 1-m  & B2         \\
51485.666& 1999.11.03& 5670-7340 &    4    &  4   & 1-m  & B2         \\
51488.416& 1999.11.06& 5670-7320 &    4    &  6   & 1-m  & B2         \\
51488.447& 1999.11.06& 4100-5780 &    4    &  6   & 1-m  & B2         \\
51514.288& 1999.12.01& 4100-5780 &    4    & 10   & 1-m  & B2,B3       \\
51635.187& 2000.03.31& 5660-7330 &    4    & 10   & 1-m  & B2          \\
51810.552& 2000.09.23& 5600-7280 &    4    &  5   & 1-m  & V2          \\
51812.555& 2000.09.25& 5600-7230 &    4    & 10   & 1-m  & S          \\
51844.500& 2000.10.27& 4100-5800 &    4    & 11   & 1-m  & B2          \\
51846.430& 2000.10.29& 5600-7240 &    4    &  5   & 1-m  & V2         \\
51848.416& 2000.10.31& 5600-7300 &    4    &  6   & 1-m  & B2          \\
51864.375& 2000.11.16& 5660-7320 &    4    &  9   & 1-m  & B2          \\
51864.448& 2000.11.16& 4100-5780 &    4    & 10   & 1-m  & B2          \\
51880.312& 2000.12.01& 5660-7330 &    4    & 17   & 1-m  & B2          \\
51885.319& 2000.12.06& 4500-7000 &    5    &164*) & 6-m  & V1,M,F    \\
51936.243& 2001.01.26& 5660-7320 &    4    & 10   & 1-m  & B2          \\
51937.234& 2001.01.27& 5660-7320 &    4    &  4   & 1-m  & V2          \\
\hline
\end{tabular}
\end{center}

\ *) aperture spectra through cirri.
\begin{center}
\begin{tabular}{lll}
Observers:\\
B1 -- Barsukova E.A. & M -- Monin D.N.       & V1 -- Valyavin G.G. \\
B2 -- Borisov N.V.   & P -- Pustilnik S.A.   & V2 -- Vlasyuk V.V.  \\
B3 -- Burenkov A.N.  & S -- Serafimovich N.I.& U  -- Ugryumov A.V. \\
F  -- Fabrika S.N.
\end{tabular}
\end{center}

Telescopes:\\
6-m -- 6-meter reflector BTA;\\
1-m -- 1-m Zeiss reflector SAO RAN.

\begin{center} Table 3.\\
The behaviour of emission lines in the outburst.
\end{center}
\begin{center}
\begin{tabular}{llrrcl}
\hline
Line  & $\lambda_{lab.}$ & $EW_f/EW_q$ & $log(F_f/F_q)$ & $log(P_i+P_e)(EV)$ & Type    \\
\hline
H$_{\gamma}$& 4340.468 &   4.86 & 0.96& 1.12& $H_{\alpha}$   \\
FeII        & 4414.78  &   2.97 & 0.80& 1.13& $H_{\alpha}$   \\
HeI         & 4471.688 &  12.28 & 1.43& 1.36& $HeI$          \\
FeII        & 4583.290 &   2.89 & 0.88& 1.13& $H_{\alpha}$?  \\
HeII        & 4685.810 & 132.82 & 2.49& 1.88& $HeII$         \\
H$_{\beta}$ & 4861.332 &   3.24 & 0.87& 1.11& $H_{\alpha}$   \\
HeI(+FeII)  & 4921.929 &   3.79 & 1.51& 1.36& $HeI$          \\
HeI         & 5015.675 &  22.90 & 1.68& 1.36& $HeI$          \\
FeII        & 5197.569 &   3.49 & 0.92& 1.12& $H_{\alpha}$   \\
FeII        & 5234.620 &   5.52 & 1.09& 1.13& $H_{\alpha}$   \\
FeII        & 5316.609 &   2.90 & 0.87& 1.13& $H_{\alpha}$   \\
FeII        & 5534.860 &   2.70 & 0.83& 1.13& $H_{\alpha}$   \\
$[NII]$     & 5754.640 &   0.32 &-0.10& 1.25& $[NII]$        \\
HeI         & 5875.792 &   8.70 & 1.35& 1.36& $HeI$          \\
FeII        & 6147.735 &   2.60 & 0.98& 1.14& $SiII$         \\
FeII        & 6318.000 &   1.23 & 0.53& 1.14& $HeI$          \\
SiII        & 6347.091 &   5.52 & 1.26& 1.26& $SiII$         \\
FeII        & 6385.470 &   1.00 & 0.54& 1.11& $HeI$          \\
FeII        & 6456.376 &   1.71 & 0.88& 1.14& $H_{\alpha}$   \\
FeII        & 6491.670 &   0.94 & 0.82& 1.13& $SiII$         \\
FeII        & 6516.053 &   1.88 & 0.77& 1.10& $H_{\alpha}$   \\
H$_{\alpha}$& 6562.816 &   1.21 & 0.60& 1.08& $H_{\alpha}$   \\
HeI         & 6678.151 &   7.92 & 1.41& 1.36& $HeI$          \\
HeI         & 7065.440 &   4.60 & 1.23& 1.36& $HeI$          \\
\hline
\end{tabular}
\end{center}
}
\newpage

\begin{center}{\bf References}  \end{center}

\begin{description}
\item{\ \ 1.} Lamers H.J.G.L.M., Zickgraf F.-J., de Winter D., et al.,//
  Astron. and Astrophys. 1998. V.340. P.117.
\item{\ 2.} Merrill P.W., Humasson M.L., Burwell C.G.//Astrophys. J.
  1932. V.76, P.156.
\item{\ 3.} Merrill P.W., Burwell C.G.//Astrophys. J. 1933. V.78, P.97.
\item{\ 4.} Downes R.A. //Publ. Astron. Soc. Pacific. 1984. V.96, P.807.
\item{\ 5.} Miroshnichenko A.S. //Astron. and Astrophys. Trans. 1995. V.6, P.251.
\item{\ 6.} Bergner Yu.K., Miroshnichenko A.S., Yudin R.V., Kuratov K.S.,
  et al.//Astron. and Astrophys. Suppl. Ser. 1995. V.112. P.221.
\item{\ 7.} Miroshnichenko A.S.//Odessa Astron. Publ. 1994. V.7. P.76.
\item{\ 8.} Chkhikvadze Ya.N.// Astrofizika. 1970. V.6. P.65.
\item{\ 9.} Smith D., Remillard R., Swank J., et al.//IAU Circ. 1998.
  No. 6855.
\item{10.} Belloni T., Dieters S., van den Ancker M.E., Fender R.P. et al.//
  Astrophys. J. 1999. V.527. P.345.
\item{11.} Ueda Y., Ishida M., Inoue H. et al.//Astrophys. J. 1998.
  V.508. L167.
\item{12.} Robinson E.L., Welch W.F., Adams M.T., Cornell M.E.//IAU Circ.
  1998. No.6862.
\item{13.} Barsukova E.A., Fabrika S.N., Pustilnik S.A., Ugryumov A.V.//
  Bull. Spec. Astrophys. Obs. 1998, V.45, P.145.
\item{14.} Clark J.S., Steele I.A., Fender R.P., Coe M.J.//
  Astron. and Astrophys. 1999. V.348. P.888.
\item{15.} Barsukova E.A., Fabrika S.N.// Variable star as a clue to
  understanding the structure and evolution of Galaxy. 2000. Ed. N.N.Samus
  and A.V.Mironov, Cygnus. Nizhny Arkhys. P.154.
\item{16.} Clark J.S., Miroshnichenlo A.S., Larionov V.M., et al.,//
  Astron. and Astrophys.. 2000. V.356. P.50.
\item{17.} Hjellming R.M., Mioduszewski A.J.//Sky and Telescope. 1998.
  V.96. No.2. P.22.
\item{18.} Hjellming R.M., Mioduszewski A.J.//IAU Circ. 1998. No.6872.
\item{19.} Orlandini M., Parmar A.N., Frontera F., et al.//Astron. and
  Astrophys. 2000. V.356. P.163.
\item{20.} Parmar A.N., Belloni T., Orlandini M., et al.//
  Astron. and Astrophys. 2000. V.360. L.31.
\item{21.} Kornilov V.G., Volkov I.M., Zakharov A.I., et al.//
  Sternberg Institute Trans. 1991. V.63.
\item{22.} Garsia M.R., Berlind P., Barton E., McClintock J.E.//
  IAU Circ. 1998. No.6865.
\item{23.} Hynes R.I., Roche P., Haswell C.A., et al.//IAU Circ. 1998.
  No.6871.
\item{24.} Deeming T.J.// Astrophys. Space Sci. 1975. V.36. P.173.
\item{25.} Kato T., Uemura M.//Inform. Bull. Var. Stars. 2001. No.5081.
\item{26.} Chentsov E.L., Klochkova V.G., Tavolganskaya N.S.//
  Bull. Spec. Astrophys. Obs. 1999. V.48. P.21.
\item{27.} Striganov A.R., Sventitskii N.S.// Tables of spectral lines of
  the neutral and ionized atoms. Atomizdat. M. 1966.
\item{28.} Heung S., Aller L.H., Feibelman W.A. et al.//Monthly Notices
  Roy. Astron. Soc. 2000. V.318. P.77.
\item{29.} Meinel A.B., Aveni A.F., Stockton M.W.//Catalog of emission lines
  in astrophysical objects. Ed.II. 1969. Tucson, Univ. of Arizona.
\item{30.} Iben I.//Astrophys. J. 1982. V.259. P.244.

\end{description}

\end{document}